\pdfoutput=1
\documentclass[fleqn,usenatbib,usedcolumn]{mnras}

\usepackage{newtxtext,newtxmath}
\usepackage[british]{babel}             
\usepackage{txfonts}                    
\let\la=\lesssim 
\usepackage[T1]{fontenc}                
\usepackage{graphicx}                   
\hypersetup{pdfauthor={C. Zurita},
pdftitle={The Historical X-ray Transient KY TrA in quiescence},
pdfkeywords={stars, KY TrA},
bookmarksnumbered=true}
\volume{{\rm in press}}

\usepackage{graphicx}	
\usepackage{amsmath}	
\usepackage{amssymb}	

\title[KY TrA in quiescence]{The Historical X-ray Transient KY TrA in quiescence}

\author[C. Zurita et al.]{
C. Zurita,$^{1,2}$\thanks{E-mail: czurita@iac.es (CZ)}
J.~M. Corral-Santana,$^{3}$
J. Casares$^{1,2}$
\\
$^{1}$Instituto de Astrof\'{\i}sica de Canarias, C/V\'{\i}a L\'actea s/n,
38200 La Laguna, Spain\\
$^{2}$Departamento de Astrof\'{\i}sica, Universidad de La Laguna, 38205 La
Laguna, Tenerife, Spain\\
$^{3}$Instituto de Astrof\'{\i}sica, Pontificia Universidad Católica de Chile
, Av. Vicu\~na-Mackenna 4860,  Macul 7820436, Santiago, Chile
}

\date{Accepted XXX. Received YYY; in original form ZZZ}

\pubyear{2015}

\begin{document}
\label{firstpage}
\pagerange{\pageref{firstpage}--\pageref{lastpage}}
\maketitle

\begin{abstract}
  We present  deep optical  images of  the {\em  historical} X-ray
  Transient  KY   TrA  in  quiescence   from  which  we   confirm  the
  identification of  the counterpart reported by  \citet{murdin77} and
  derive   an   improved    position   of   $\alpha$=15:28:16.97   and
  $\delta$=-61:52:57.8.  In  2007 June  we obtained  $I$, $R$  and $V$
  images,  where the  counterpart seems  to be  double indicating  the
  presence of an interloper  at $\sim$1.4\,arcsec NW. After separating
  the  contribution   of  KY  TrA  we   calculate  $I$=21.47$\pm$0.09,
  $R$=22.3$\pm$0.1  and $V$=23.6$\pm$0.1.   Similar brightness  in the
  $I$ band  was measured in May  2004 and June 2010.   Variability was
  analyzed from series  of images taken in 2004,  spanning 0.6\,h, and
  in two blocks of  6\,h during 2007.  We find that  the target is not
  variable  in any  dataset above  the error  levels $\sim$0.07\,mags.
  The presence  of the interloper  might explain the  non-detection of
  the  classic  ellipsoidal modulation;  our  data  indicates that  it
  contributes  around half  of  the  total flux,  which  would make  a
  variation <0.15 mags not detectable.   A single spectrum obtained in
  2004  May  shows  the  $H\alpha$ emission  characteristic  of  X-ray
  transients    in   quiescence    with   a    full-width-half-maximum
  $FWHM=2700\pm$ 280 km  s$^{-1}$.  If the system follows  the FWHM --
  K$_2$ correlation found by  \citet{casares15}, this would correspond
  to   a    velocity   semi-amplitude    of   the   donor    star   of
  K$_2$=630$\pm$74\,km s$^{-1}$.  Based on  the outburst amplitude and
  colours of the  optical counterpart in quiescence we  derive a crude
  estimate of the orbital period of 8\,  h and an upper limit of 15\,
  h  which  would  lead  to   mass  function  estimates  of  $\approx$
  9\,M$_{\odot}$ and $<$16\,M$_{\odot}$ respectively.
\end{abstract}




\section{Introduction}

Low-mass  X-ray  binaries  (LMXBs)  are  mass-exchange  binaries  that
contain an accreting black hole or neutron star primary and a low-mass
secondary star.  Accretion takes place through an accretion disc which
encircles the compact  object and regulates the flow  of material onto
it making these objects the brightest  X-ray sources in the sky. LMXBs
provide  an ideal  playground  for exploring  the  physics of  compact
objects yielding  the confirmation  of the  existence of  stellar mass
black  holes.  Optical  observations are  crucial to  prove this.   By
observing the  radial velocity  curve of the  companion star,  one can
determine the mass function of  the system, which represents a minimum
mass for  the accreting compact  object.  This experiment can  be best
applied to a subclass of LMXBs,  the so called X-ray Transients (XTs),
in  which  X-ray activity  occurs  only  during well-defined  outburst
episodes.   Between outbursts,  the emission  from the  accretion flow
fades to the  point that the companion star is  clearly visible and it
is nearly undisturbed by irradiation; hence it can be used to derive a
dynamical mass for the compact object \citep[eg.,][]{charles06}.\\

The distribution of black hole masses  can only be determined from the
study   of  X-ray   binaries  \citep[eg.,][]{casares07}   and  it   is
intricately related to the population  and evolution of massive stars,
the energetics and dynamics of  supernova explosions, and the critical
mass  dividing neutron  stars and  black holes.   Several attempts  to
extract statistical  information from  the observed  mass distribution
have  been  made \citep{bailyn98,ozel10,farr11,kreidberg12},  however,
the  small number  of black  holes  prevents us  from extracting  very
compelling conclusions.  About $\sim$10$^8$-10$^9$ stellar--mass
black holes are believed to  exist in the Galaxy \citep{brown94} while
$\sim$10$^3$--10$^4$   are    expected   to   be   members    of   XTs
\citep{white96,romani98,yungelson06}.   Unfortunately   only  $\sim$20
black hole  candidates have reliable dynamical  mass determinations to
date. In summary, it becomes necessary to increase the sample of black
hole masses.\\

The best place to look for black  holes are XTs, which are detected in
outburst by  X-ray all-sky monitors.  This  was the case of  KY TrA, a
{\em historical}  X-ray transient  discovered in 1974  by the  Ariel V
instruments \citep{pounds74}.  After a short precursor, KY TrA reached
an  outburst peak  flux  of  0.9\,Crab in  the  3-6\,keV band,  before
decaying    with   an    e-folding    time   of    about   2    months
\citep{kaluziensky75}.   The  source  also   showed  a  low  intensity
outburst in 1990 which was  significantly fainter than the discovery
1974 outburst  \citep{barret95}. Six months  later, an upper  limit to
the          quiescent           X-ray          luminosity          of
$\sim$2-10$\times$10$^{33}$erg\,s$^{-1}$  was derived.   The ultrasoft
X-ray spectrum  seen by Ariel  V and the  hard tail observed  by SIGMA
\citep{barret92}  strongly  suggests  that  KY TrA  is  a  black  hole
candidate.  The optical  counterpart was identified 12  days after the
X-ray  maximum  of the  1974  outburst  at $B$=17.5  \citep{murdin77}.
Surprisingly, no further optical studies have been done since.\\

In this  paper we present an  optical study of the  counterpart of the
X-ray Transient  KY TrA in  quiescence to test its  identification and
obtain information about the donor star.

\section{Observations}

\subsection{Photometry}

The field of KY TrA was observed on UT 2004 May 16 with the ESO VLT U4
telescope  at  Paranal  and  the  FOcal  Reducer  and  low  dispersion
Spectrograph (FORS2). Seven 300\,s  exposures were obtained in Johnson
$I$  , spinning  over 0.6\,hours.  We also  obtained series  of 600\,s
images in $I$ on  UT 2007 Jun 16 and 17 on the  3.6\,m telescope at La
Silla Observatory, equipped with the ESO Faint Object Spectrograph and
Camera (EFOSC). In  the first night of this run  the sky was partially
covered by clouds and no calibrations  or images in other filters than
$I$  were obtained.   Fortunately weather  improved on  Jun 17  and we
could take also $V$ and $R$ images  of KY TrA with 1500\,s and 1000\,s
respectively. Furthermore $R$, $I$ and  $H\alpha$ images were taken on
UT 2010 June 03 on the ESO VLT U1 telescope at Paranal and FORS2 using
80\,s, 45\,s and 400\,s integration respectively in each band.\\

For calibration,  we observed the  field of KY TrA  on UT 2015  Jan 18
with the 1.3\,m SMARTS telescope at Cerro Tololo (Chile). We took $V$,
$R$ and  $I$ images with,  respectively, 800,  500 and 500  seconds of
exposure  time. We  observed  also  the Landolt  field  SA107 with  30
seconds integration  in each of the  three bands. An observing  log is
presented in Table 1.\\

\begin{table*}
 \centering
  \caption{Log of the observations.}
  \begin{tabular}{lcccc}
{\bf Photometry:} & & & & \\
  \hline    
  Date & Exp. time (s) & $\Delta$T (hr) & Filter &  Telescope /\\
       &               &                &         & Instrument \\
  \hline
2004/05/16 & 300 & 0.6 & $I$ & ESO VLT U4 / FORS2 \\
2007/06/16 & 600 & 6.3 & $I$ & ESO 3.6\,m / EFOSC  \\
2007/06/17 & 600 & 6.0 & $I$ &  '' \\
      ''   & 1000 & - & $R$  &  ''\\
      ''   & 1500 & - & $V$  &  ''\\
2010/06/03 & 80   & - & $I$  & ESO VLT U1 / FORS2 \\
      ''   & 45   & - & $R$  &  ''\\
      ''   & 300  & - & $H\alpha$ &  ''\\
2015/01/18 & 500   & - & $I$  & SMARTS 1.3\,m / ANDICAM \\
      ''   & 500   & - & $R$  &  ''\\
      ''   & 800   & - & $V$  &  ''\\
\\
{\bf Spectroscopy:} & & & & \\
    \hline
  Date & Exp. time (s) &  & Grism &  Telescope /\\
       &               &                &         & Instrument \\
  \hline
    2004/05/16 & 2000 &  & GRIS-600RI & ESO VLT U4 / FORS2 \\
    \hline
\end{tabular}
\end{table*}

All images  were corrected for  bias and flat-fielded in  the standard
way  using {\sc  IRAF}. In  particular the  2007 $I$-band  frames were
affected  by fringing  so these  images  were corrected  using a  {\em
  master flat},  i.e.  the result of  the median of a  set of dithered
exposures of our field. Point spread function {\sc PSF} photometry was
obtained with {\sc DAOPHOT  II} \citep{stetson87} whereas differential
photometry  of the  2004 and  2007  series were  performed creating  a
photometric  reference level,  through an  {\em ensemble}  of isolated
stars  in the  KY  TrA  field, following  the  technique described  in
\citet{honeycutt92}.\\

We finally calibrated  a set of stars  in the 2015 field  that we then
used as  secondary standards  to calibrate the  2004 and  2007 images,
taken  under  non photometric  conditions.   To  do  so, we  used  the
observatory  extinction coefficients  together  with  the zero  points
extracted from the Landolt standards.

\subsection{Spectroscopy}

A single  2000\,s spectrum was  obtained on the  night of 2004  May 16
using  FORS2 at  the ESO  VLT U4  telescope at  Paranal.  We  used the
GRIS-600RI grism,  and a  2$\times$2 binning in  both the  spatial and
spectral   direction,  which   provides  a   wavelength  coverage   of
$\lambda\lambda$5200--8400  at  1.68\,\AA\,pix$^{-1}$ dispersion  and
600 km s$^{-1}$ resolution.  Standard  procedures were used to de-bias
and  flat-field   the  spectra.   The  one-dimensional   spectra  were
extracted using  optimal extraction routines which  maximize the final
signal-to-noise ratio.  A Hg--Cd--Ar--Ne  arc was obtained  to provide
the wavelength calibration scale.

\section{Astrometry of KY TrA}
\label{astrometry}

\begin{figure}
\includegraphics[width = \textwidth,bb=24 0 818 596]{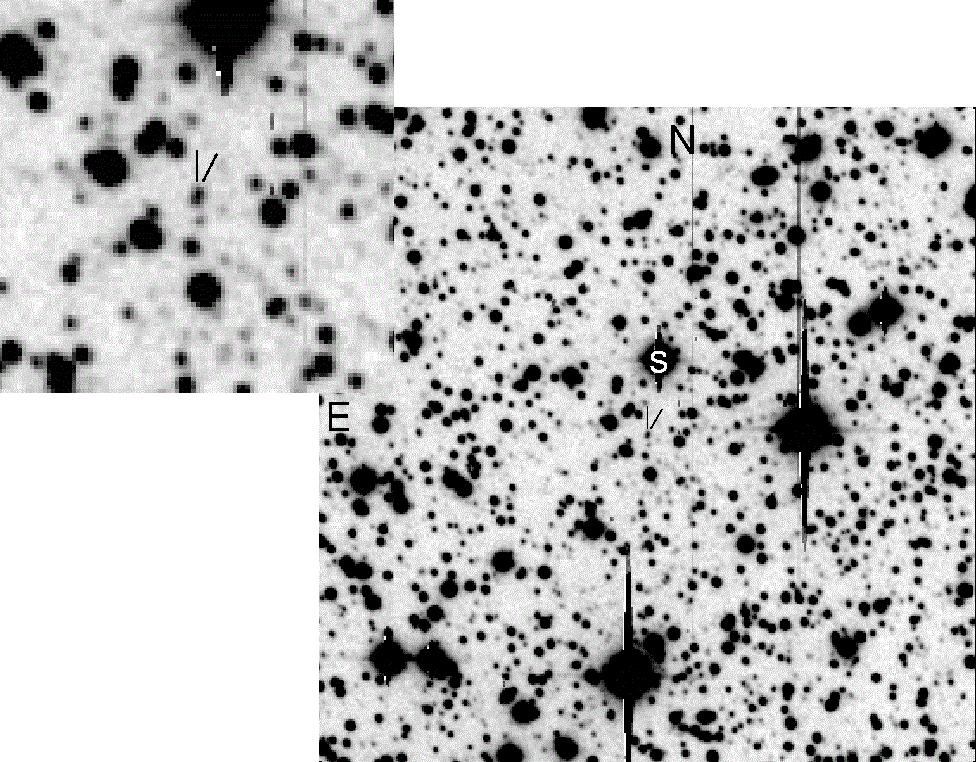}
\caption{A 1800\,s $I$ image of KY TrA  taken in 2007, 17 June, with a
  field of  view of 2x2\,arcmin  and a zoom  of the central  region of
  30x30\,arcsec. North is at the top, and  East is to the left and the
  plate scale  is 0.315  arcsec per  pixel.  Star  labeling as  'S' in
  \citet{murdin77} is also  shown for comparison.  The  target, in the
  middle  of the  fields,  is  marked.  An  elongation  of the  source
  profile along  the NW direction  is clearly visible  indicating that
  the counterpart is double and it is blended with an interloper.  The
  XT   is  the   northern   component  of   the   pair  (see   section
  \ref{astrometry}).  }
\label{figure:field}
\end{figure}

The   optical    counterpart   of    KY   TrA   was    identified   at
$\alpha$=15:28:16.59     and     $\delta$=-61:52:58.1    (2000)     by
\citet{murdin77} 12  days after the  outburst. These authors  showed a
Schmidt plate of  the proposed counterpart but no deep  image with the
target  in  quiescence   has  been  published  to   date.   In  Figure
\ref{figure:field}  we  show  an  improved finding  chart  of  KY  TrA
($I$-band, 1800\,s exposure and 0$.\!\!^{\prime\prime}$9 seeing, taken
on  2007 June  17).  The  field  is 2x2\,arcmin  and the  star at  the
position proposed  by \citet{murdin77}  is marked  in the  center. For
comparison we have also marked the  star labeled as 'S' in that paper.
In  this  figure  we  also  show  a zoom  of  the  central  region  of
30x30\,arcsecs where an elongation of  the source profile along the NW
direction is  clearly visible.   We measure an  elongation coefficient
(the  IRAF ``ellipsoidal''  parameter) of  0.68, significantly  larger
than  the  typical values  obtained  for  nearby field  stars  (0.32).
Although  this requires  confirmation  through  better seeing  quality
images, it  strongly suggests that  the counterpart is double  and the
transient is blended with an interloper.\\

To obtain a precise astrometric solution, we used the positions of the
astrometric   standards   selected   from  the   USNO-B1   astrometric
catalog\footnote  {USNO-B1 is  currently incorporated  into the  Naval
  Observatory  Merged  Astrometric  Data-set  (NOMAD)  which  combines
  astrometric  and photometric  information  from Hipparcos,  Tycho-2,
  UCAC,      Yellow-Blue6,      USNO-B,       and      the      2MASS,
  www.nofs.navy.mil/data/fchpix/}        with         a        nominal
0$.\!\!^{\prime\prime}$2  uncertainty.   Hundreds of reference
objects can  be identified in  our field  from which we  selected 392,
discarding the stars with significant  proper motions.  The IRAF tasks
ccmap/cctran were  applied for  the astrometric transformation  of the
images.   Formal rms  uncertainties  of the  astrometric  fit for  our
images are $\la$0$.\!\!^{\prime\prime}$25 in  both right ascension and
declination,  which is  compatible with  the maximum  catalog position
uncertainty  of   the  selected   standards.   The  star   within  the
\citet{murdin77}  error box  has coordinates  $\alpha$=15:28:16.97 and
$\delta$=-61:52:58.2 with  a conservative  estimate of our  3$\sigma $
astrometric uncertainty  of $\la$0 $.\!\!^{\prime\prime}$3 in  both RA
and Dec.\\

\begin{figure}
\includegraphics[width=90mm]{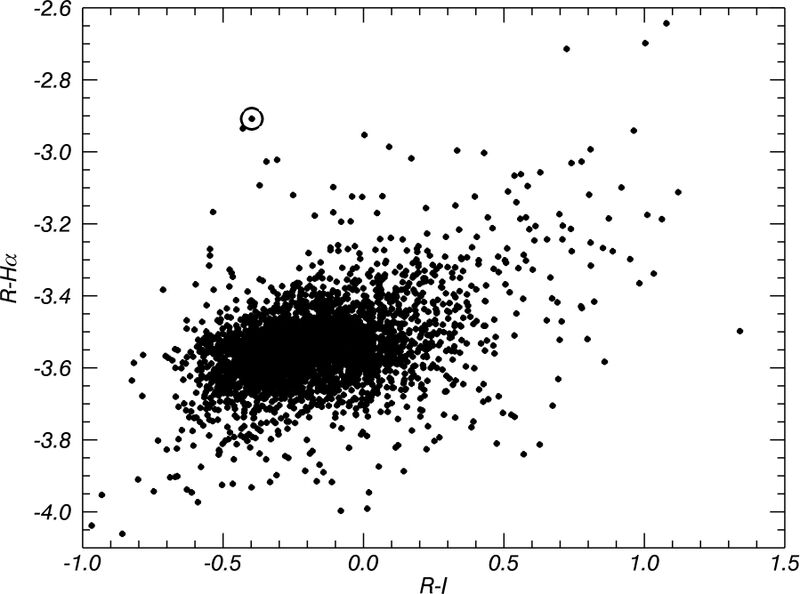}
\caption{KY TrA uncalibrated ($R$-$I$)--($R$-$H\alpha$) diagram of all
  the objects detected in the three  photometric bands in our field of
  view.   Our proposed  target, marked  with a  circle, shows  a clear
  $H\alpha$ excess above the main  stellar locus. Under the hypothesis
  that the light of KY TrA is contaminated by an interloper, it should
  be  located even  higher in  this diagram,  with a  larger $H\alpha$
  excess.  The source closest to KY TrA is unknown.}
 \label{figure:diagram}
\end{figure}

We  test  this identification  by  cross-matching  the $R$,  $I$,  and
$H\alpha$ photometry of the field of KY TrA taken on 2010 to build the
($R$-$I$)--($R$-$H\alpha$) diagram of all  the objects detected in the
three photometric bands. Our proposed  target (marked with a circle in
Figure \ref{figure:diagram}) shows a  clear $H\alpha$ excess above the
main stellar locus, confirming it is the true quiescent counterpart of
KY  TrA.  Note  that very  close to  the KT  TrA counterpart  there is
another source of similar colour  and H$\alpha$ excess. This source is
at 58\,arcsec from  the XT and its nature is  unknown. However, KY TrA
should be located even higher in this diagram since the presence
of the interloper dilutes its actual H$\alpha$ excess.\\

\begin{figure}
\includegraphics[width=90mm]{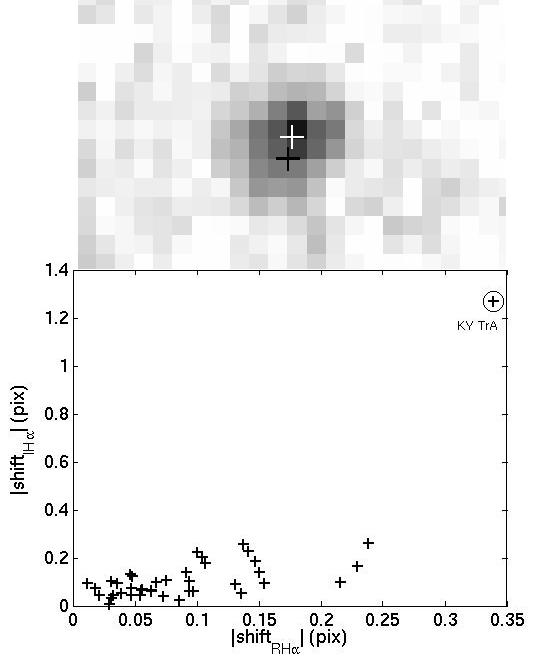}
\caption{ (Upper) Zoom of the $H\alpha$  profile of KY TrA with a white
cross  marking  its  centroid.  The black  cross  marks  the  centroid
measured in the $I$ image.  (Lower)  Modulus of the shifts (in pixels)
between the  image centroid positions  for KY TrA and  ~40 surrounding
stars as  measured in our $R$,  $I$ and $H\alpha$ images  (for details
see text).}
 \label{figure:centroids}
\end{figure}

 We can  refine the location of  the counterpart by choosing  which of
 the components of the blend is actually the XT counterpart. To do so,
 we  first  aligned  the  $R$,  $I$, and  $H\alpha$  images  and  then
 calculated the centroids  of the profile targeted as KY  TrA and of a
 set of 40 stars around it.  In Figure \ref{figure:centroids} (bottom)
 we show  the modulus of  the shifts  between $R$, $I$  and $H\alpha$,
 defined     as      $[(x_R-x_{Ha})^2+(y_R-y_{Ha})^2]^{(1/2)}$     and
 $[(x_I-x_{Ha})^2+(y_I-y_{Ha})^2]^{(1/2)}$   where  $(x,y)$   are  the
 positions  of  the centroids  in  pixels.   The  centroid of  KY  TrA
 measured in H$\alpha$ is clearly shifted with respect to its position
 in   the   $R$  and   $I$   images.    The   top  panel   in   Figure
 \ref{figure:centroids} shows  a zoom of the  H$\alpha$ image centered
 on our  target, with  the white  cross marking  its centroid  and the
 black  cross  the centroid  measured  in  the $I$-band  image.   This
 indicates that  the XT is the  component of the blend  located at the
 NW.  Assuming that  the $H\alpha$ centroid is the  actual position of
 the   target,   we   derived   an   improved   source   position   of
 $\alpha$=15:28:16.97 and $\delta$=-61:52:57.8  with an uncertainty of
 $\la$0 $.\!\!^{\prime\prime}$3 in both RA and Dec.

\onecolumn
\twocolumn

\section{Photometry}

\subsection{The colours of KY TrA}

The colours of KY TrA in quiescence  were obtained on UT 2007 June 17.
We     calculated    $I$=20.88$\pm$0.01,     $R$=21.75$\pm$0.01    and
V=22.83$\pm$0.12.  Photometric  error estimates on the  magnitudes are
based  on   a  combination  of   Poisson  statistics  and   the  error
contribution  of the  stars used  for calibration.   These magnitudes,
however, correspond to the transient  blended with the interloper (see
section \ref{astrometry})  so they  need to  be corrected  taking this
into  account.   To  do  so,   we  cleaned  the  contribution  of  the
contaminant component  by subtracting its  best PSF fit.   The initial
centers for fitting the profiles  of the components were determined by
visual inspection  of the $I$ image,  where they are more  clear.  The
two components  of the blend  have approximately equal  brightness: we
calculated  $I$=21.47$\pm$0.09, $R$=22.3$\pm$0.1  and V=  23.6$\pm$0.1
for the top component, which is the XT counterpart as suggested by the
$H\alpha$ image. The  $(V-R)$ and $(V-I)$ colours are typical  of a M0
star,  although  note that  this  is  not corrected  for  interstellar
reddening.  \citet{murdin77}  report a reddening  E(B-V)$\geq$0.5.  An
estimation of  the interstellar  reddening for any  sky region  can be
obtained   from  the   NASA/IPAC  Infrared   Science  Archive\footnote
{http://irsa.ipac.caltech.edu/applications/DUST/}.     The   reddening
quoted  for the  field  of  KY TrA,  E(B-V)=0.7,  implies a  corrected
$(V-R)_0$=0.77  and  $(V-I)_0$=1.15   consistent  with  a  $\sim$K0-2V
companion. Nevertheless,  it should  be noted that  this is  likely an
upper limit  to the true  spectral type of  the donor star  because we
have neglected  any contribution from  a residual accretion  disc into
the observed colour.

\subsection{Searching for variability}

\begin{figure}
\includegraphics[width=90mm, bb=0 0 420 582]{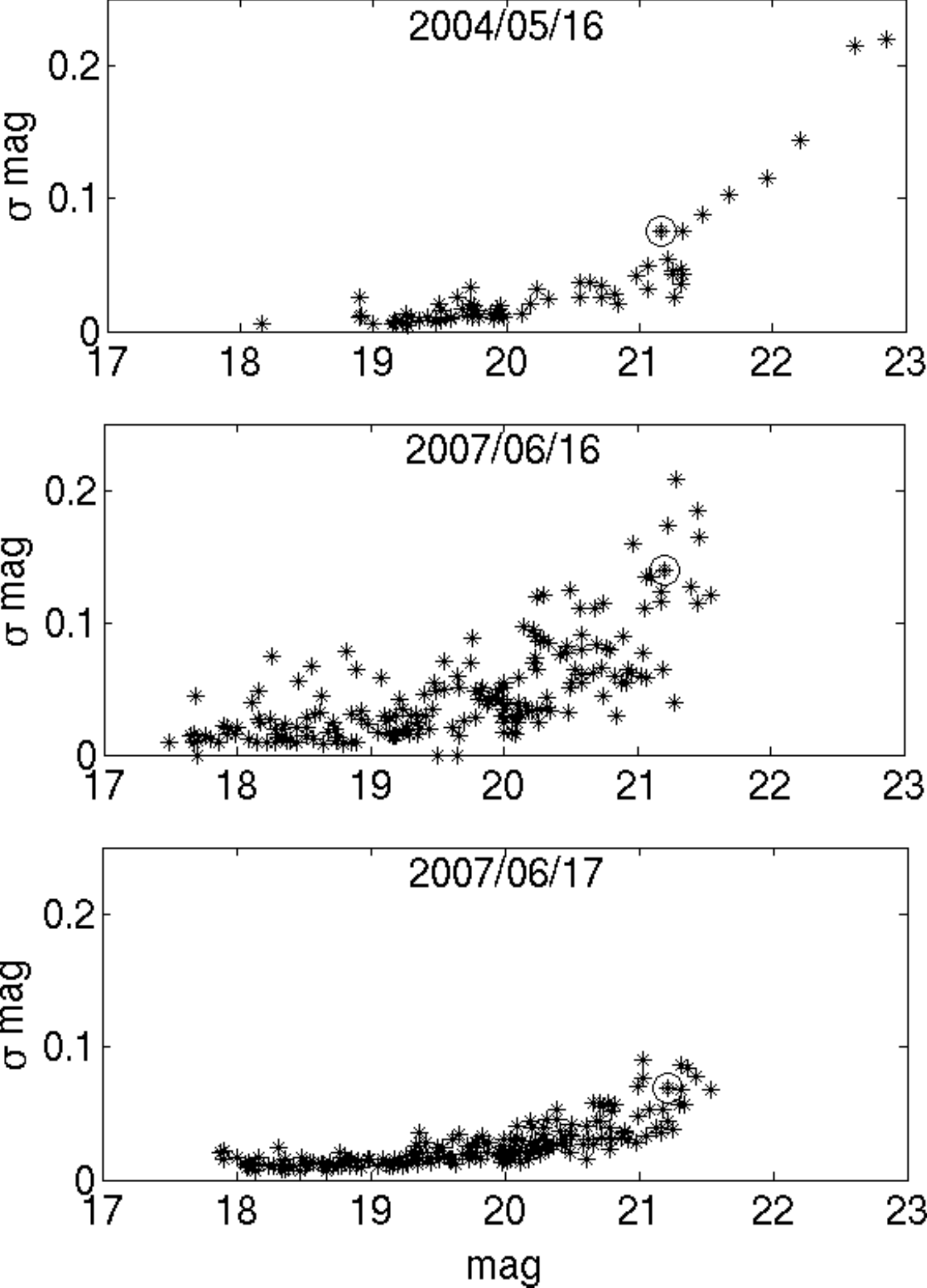}
\caption{Scatter of each individual star  from its mean magnitude from
  the $I$ band series of images of  each night. The blend (i.e, KY TrA
  plus the interloper) is marked with a circle.}
 \label{figure:variability}
\end{figure}

The limited seeing conditions prevents  us from deblending the flux of
the  two stars  in  every  single image.   Therefore,  we studied  the
variability of KY TrA by integrating  the total flux of the blend from
the series of  images in the $I$-band in the  two different epochs: in
2004 May with less  than one hour of observations and  in 2007 June 16
and 17  with about 6  hours coverage.   The 2004 images  were obtained
during twilight so  photometry accuracy was dominated by  a bright sky
level. Unfortunately on 2007 June 16  we were affected by poor weather
conditions with variable transparency caused  by clouds.  On 2007 June
17 no clouds were present although sky transparency was not ideal.  In
this night we get $\sim$7\% photometry  for a 21.5 magnitude star.  In
summary, we  found that  the target  is not  variable above  the error
levels in any of our  nights (i.e., $\sigma$=0.07, 0.14 and 0.07\,mags
for 2004 May  and 2007 June 16 and 17  respectively).  This is clearly
seen in Fig.\ref{figure:variability} where we  plot the scatter in the
observed magnitudes.  The scatter around  the mean magnitude of KY TrA
(marked with  a circle in Fig.\ref{figure:variability})  is consistent
with that  displayed by  the field stars  of similar  mean brightness.
Note  that  the  presence  of   the  interloper  dilutes  the  orbital
modulation  of KY  TrA, which  would explain  the lack  of photometric
variability on our images.   Because the interloper contributes around
half of the  total flux, we can  only conclude that the  target is not
variable by $>$0.15\,mags. \\

After re-scaling the zero-point of each  night we found that the total
flux remains  stable at $I$=20.88$\pm$0.01 ($I$=21.47$\pm$0.09  for KY
TrA, i.e.  the top component of the blend) in our entire dataset, from
2004 until 2007.\\
 
\begin{figure}
\includegraphics[width=90mm]{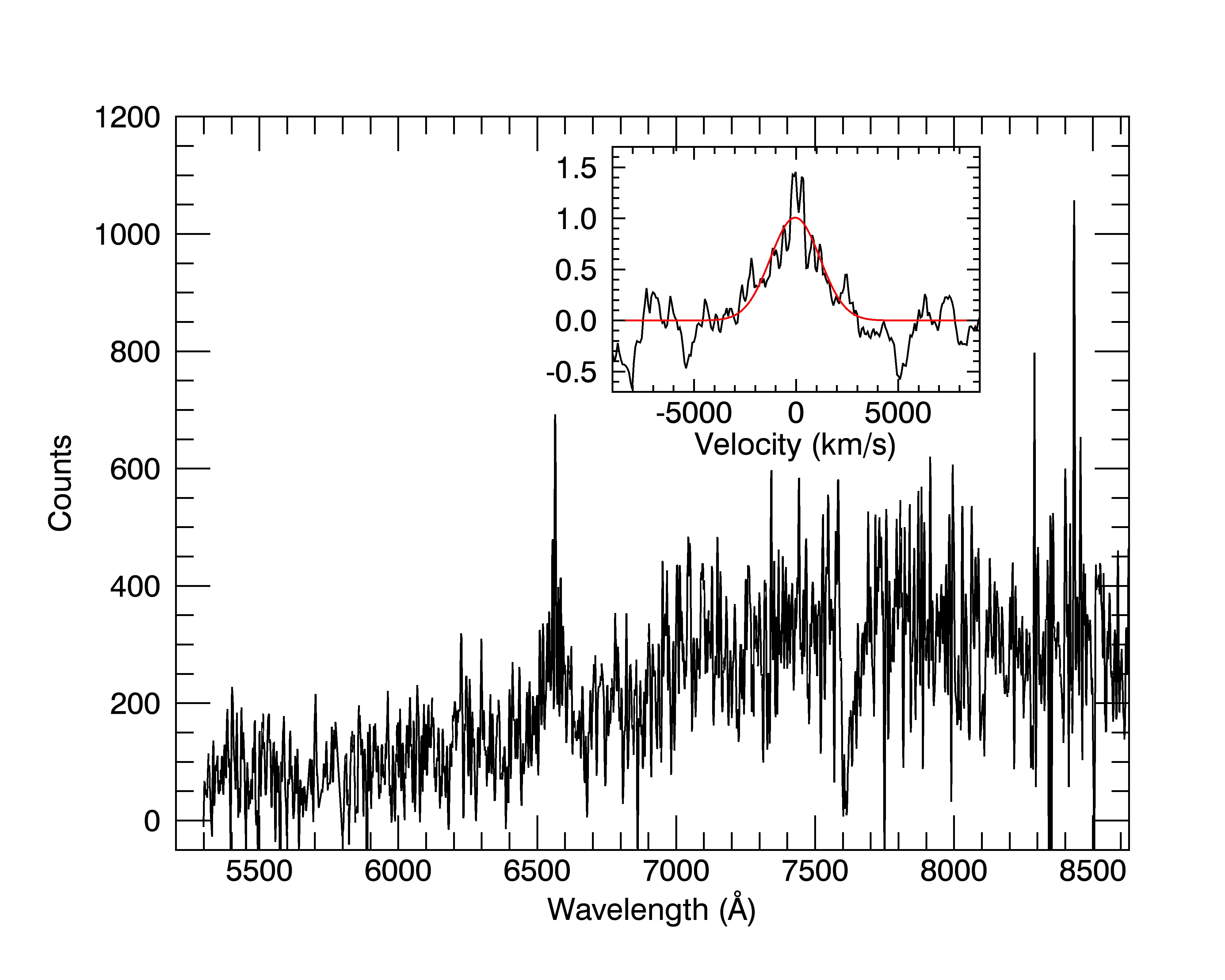}
\caption{ VLT optical spectrum of KY TrA from 2004 May 16, and (inset)
  a Gaussian fit to the H$\alpha$ profile.}
 \label{figure:Ha_gaussian}
\end{figure}

\section{Spectroscopy}

Although very  noisy, our spectrum  shows the H$\alpha$  emission line
characteristic  of X-ray  transients in  quiescence.  We  obtained its
full-width-half-maximum (FWHM)  from a  Gaussian fit of  the continuum
rectified spectrum within the  range $\pm$10000\,km/s, centered on the
H$\alpha$  line.   This  is  shown  in  Fig.~\ref{figure:Ha_gaussian}.
After subtracting  quadratically the  instrumental resolution  we find
FWHM=2700$\pm$280\,km/s where  the error  is the formal  1-$\sigma$ on
the fitted parameter as derived through $\chi^2$ minimization. We note
that the  quoted error is  within the typical 10\%  standard deviation
caused by  intrinsic line  variability and therefore  we take  this as
realistic \citep{casares15}.   We also extracted the  equivalent width
(EW) by  integrating the H$\alpha$ flux  after continuum normalization
and find $EW=72\pm7$\AA. Note, however,  that this value is diluted by
the extra continuum of the interloper and  hence the true EW of KY TrA
is underestimated by a factor ~2.

\section{Discussion}

Although KY TrA  is a very promising black hole  candidate, it had not
been studied in the optical band  since its discovery in 1974. We have
observed KY TrA  in quiescence and confirmed  its identification.  The
finding  chart we  present  will certainly  be  helpful in  performing
observations with eELT-class telescopes  in order to obtain dynamical
information on the mass of the compact object.\\


A  rough  estimate  of  the  period  can  be  made  by  combining  the
\citet{paczynski71} expression for the averaged radius of a Roche lobe
with Kepler's Third Law to get the well-known relationship between the
secondary's     mean    density     and     the    orbital     period:
$\rho=(110/P_h^2)$\,g\,cm$^{-3}$, where $\rho$ is the mean density and
$P_h$  the orbital  period in  hours.  Under  the hypothesis  that the
light of  KY TrA is  contaminated by  the interloper, we  calculate an
orbital period of about 8\,h assuming  a K0V star.  It should be noted
that this is likely  an upper limit since, as we  pointed out in Sect.
4.1, we  have neglected  any residual  contribution from  an accretion
disc to the  colour of KY TrA.  An independent  estimation can be made
using  the empirical  relation $\Delta\,V=14.36-7.63\,log\,P_h$  which
predicts  the orbital  period of  XTs with  orbital periods  less than
1\,day   given  only   its  visual   outburst  amplitude   $\Delta\,V$
\citep{shahbaz98}.  During outburst, optical  emission is dominated by
the reprocessing of the X-rays in the accretion disc where most of the
reprocessed energy is  radiated in the ultraviolet.   According to the
irradiated model predictions $(B-V)_{disc}\sim$0 \citep{vanparadijs95}
resulting  in  $V\sim$17.5  at  the  outburst  peak  \citep{murdin77}.
Taking  $V$=23.6  in  quiescence,  the  total  outburst  amplitude  is
$\sim$6.1\,mag  so this  would  lead  to an  orbital  period of  about
12\,hours.  We can place a robust  upper limit for the period of about
15\,h corresponding to a minimum outburst amplitude if the source were
not contaminated by any interloper. \\

After  analyzing the  H$\alpha$  emission profiles  of 12  dynamically
confirmed black  holes and  2 neutron star  X-ray transients  (XTs) in
quiescence, \citet{casares15}  has found  a tight  correlation between
the FWHM of the H$\alpha$ line  and the velocity semi-amplitude of the
donor star,  where K$_2$=0.233(13)$\times$FWHM.  We have  applied this
relation to  KY TrA and  predict K$_2$=630$\pm$74\,km/s.  This  can be
combined with our  rough estimates of the orbital period  to infer the
mass function $f(M)$ of the  binary. Our upper limit $P<15$\,h implies
$f(M)<16$   M$_{\odot}$    while   $P\approx8$\,h   would    lead   to
$f(M)\approx9$  M$_{\odot}$.   More  accurate constraints  require  an
accurate determination of the orbital period.\\

Despite the foregoing,  no variability has been found  above the error
levels, i.e.  $\sim$0.07\,mags,  indicating that we may  be looking at
the binary at very low  inclination.  However, given the contaminating
flux  from an  interloper, the  variability  would be  diluted to  the
extent that any intrinsic  variability above $\sim$0.15\,mag would not
be detectable given our error levels, and so the inclination might not
be as  low.  Interestingly, KY TrA  has one of the  broadest H$\alpha$
lines among SXTs (a summary of the  parameters of the XTs can be found
in \citet{casares15} Table 3), suggesting that, if KY TrA is viewed at
low inclination, it must have a very short orbital period.  Some clues
about the  inclination are provided by  the EW of the  H$\alpha$ line,
since it  depends on the  binary geometry.   The EW tends  to increase
with inclination because, when the disc is seen at large inclinations,
its  continuum brightness  decreases.  An  interesting exercise  is to
locate  KY  TrA  in  the  EW--FWHM   diagram  shown  in  figure  5  of
\citet{casares15}.  Regions of constant inclination and $M_1/P$, where
$M_1$ is the mass of the compact object and $P$ is the orbital period,
are defined in the diagram. Both  the relatively large EW and the high
$M_1/P$ factor  are reminiscent of  XTE J1118+480 and suggest  that KY
TrA   also   hosts   a   black    hole   seen   at   moderately   high
inclination. Furthermore,  we point  out that given  the contaminating
flux from an interloper, the observed $EW$ would just be a lower limit
to  the true  $EW$ since  the latter  would be  diluted by  the excess
continuum.  Clearly,  more higher  quality photometry is  necessary to
resolve these issues and draw further firm conclusions.

\section*{Acknowledgements}

JMC-S acknowledges  financial support  {\bf from} CONICYT  through the
FONDECYT project  No.  3140310 and  Basal-CATA PFB-06/2007, and  JC to
the Spanish Ministerio de Educaci\'on, Cultura y Deportes under grants
AYA2010--18080 and AYA2013-42627.  We specially  want to thank Jose L.
Prieto for  his help  in obtaining the  SMARTS observations.   We also
acknowledge the referee (Phil Charles) whose comments greatly improved
the manuscript.













\bsp	
\label{lastpage}
\end{document}